\documentclass[]{piparticle-final}
\usepackage{graphicx}
\usepackage{amsmath}

\usepackage{cite} % To compress citation ranges
\usepackage{epstopdf} 

\begin{document}

\volume{7}               % To be inserted by Editor
\articlenumber{070002}   % To be inserted by Editor
\journalyear{2015}       % To be inserted by Editor
\editor{C. A. Condat, G. J. Sibona}   % To be inserted by Editor
%\reviewers{}  % To be inserted by Editor
\received{20 November 2014}     % To be inserted by Editor
\accepted{10 March 2015}   % To be inserted by Editor
\runningauthor{F. A. S. Ferrari \itshape{et al.}}  % To be inserted by Editor
\doi{070002}         % To be inserted by Editor

\title{Two distinct desynchronization processes caused by lesions in globally coupled neurons}

% Institution references with \cite are inserted after \maketitle in theaffiliation enviroment
\author{Fabiano A. S. Ferrari,\cite{inst1}\thanks{Email: fabianosferrari@gmail.com}\hspace{0.5em}
	Ricardo L. Viana\cite{inst1}
	}

\pipabstract{
To accomplish a task, the brain works like a synchronized neuronal network where all the involved neurons work together. When a lesion spreads in the brain, depending on its evolution, it can reach a significant portion of relevant area. As a consequence, a phase transition might occur: the neurons desynchronize and cannot perform a certain task anymore. Lesions are responsible for either disrupting the neuronal connections or, in some cases, for killing the neuron. In this work, we will use a simplified model of neuronal network to show that these two types of lesions cause different types of desynchronization.
}

\maketitle

\blfootnote{
\begin{theaffiliation}{99}
	\institution{inst1} Physics Department, Universidade Federal do Paran\'a, Curitiba, Brazil.
\end{theaffiliation}
}

%\maketitle

\section{Introduction}

The neuronal dynamics can be represented as a dynamical system and a population of neurons as a neuronal network. The mean electrical field amplitude of a population of neurons has neglected values when they are uncoupled or weakly coupled. This amplitude is enhanced when the coupling between them is high enough to make them synchronized among themselves \cite{Batista_1}. At the synchronized state, it is possible to measure the mean electrical activity of a large number of closed neurons using EEG \cite{Tatum, Nunez}. Abnormalities or absence of synchronization have been reported as a consequence of neurodegenerative diseases \cite{Stam_06,Lo}. This dynamical effect is a consequence of topological changes caused by lesions spreading in the brain. However, every disease has its own features and, here, we propose to study the different dynamical effects caused by different types of 
lesions.

Measures of neuronal functional activity using EEG \cite{Stam} and fMRI \cite{Eguiluz} have shown spatiotemporal patterns formation. An explanation for this behavior is the emergence of a critical state in the neuronal dynamics providing conditions for a formation of distinct clusters at the functional level \cite{Haimovici_13}. When the neuronal population is considered as a complex network structure with a hierarchical-modular architecture, this high heterogeneity is related to a stretching of criticality and consequently increased functionality \cite{Moretti_13}. Recent papers have shown functional differences between healthy and unhealthy patients with different neuropathologies \cite{Stam_06, Ponten, Heuvel}. Schizophrenia, for example, has been related to neuronal decoupling \cite{Cabral_12_2}. 

Unfortunately, many papers are constrained to the study of functional connections and the comparison between healthy and unhealthy patients. Efforts have been made to explain the dynamical changes caused by lesions and different models have been proposed to connect what happens in the neuronal level to what happens in the macroscopic level \cite{Haimovici_13, Cabral_12_1}. Nevertheless, a complete understanding about the dynamical effect of lesions in the brain is still missing.

From the point of view of electrical activity, when the brain needs to execute a specific task there is a group of neurons that synchronize and work together to perform it. When a lesion spreads in the brain, depending on its size, it can disrupt important connections and certain tasks cannot be done anymore. Based on this hypothesis, we present here a simplified neuronal network model of globally coupled neurons to study the effects of the desynchronization induced by a lesion spreading randomly in the brain. We focus on two main cases: one in which the connections among neurons are disrupted and a second one in which the lesion kills the neurons. Despite of the simplicity of the model, the observed phase transitions from synchronized to the desynchronized state have shown different properties for these two types of lesions. 

\section{Model}

In this work, we consider a network of Rulkov neurons globally coupled (mean field). However, other neuronal models could be used and provide similar results, for example: Kuramoto \cite{Kuramoto_75}, Hindmarsh-Rose \cite{Xia_05} and Morris-Lecar \cite{Wang_05}. Rulkov neurons are described by a fast variable $x$ and a slow variable $y$. The dynamic associated with each neuron in the network can be described as
\begin{eqnarray}
x_{n+1}^{(j)}&=&\frac{\alpha^{(j)}}{(1+(x^{(j)}_n)^2)}+y^{(j)}_n+\frac{\varepsilon}{N}\sum_{i=1}^{N}x^{(i)}_n,\label{eq1}\\
y^{(j)}_{n+1}&=&y^{(j)}_n-\sigma x^{(j)}_n-\beta,\label{eq2}
\end{eqnarray}
where $\sigma=\beta=0.001$, $\alpha^{(j)}$ is a bifurcation parameter randomly chosen in the interval $[4.1,4.3]$, exhibiting bursts, $N$ is the network size and $\varepsilon$ is the coupling strength \cite{Rulkov}. 

The first step in our analysis is to choose an appropriate coupling strength such that the network becomes synchronized. To characterize phase synchronization, we will define a geometric phase for each neuron. Considering one period of oscillation and the distance between two successive bursts,  the phase of each neuron $j$ is given as \cite{Batista_1},
\begin{eqnarray}
\varphi_n^{(j)}=2\pi k+2\pi \frac{n-n_k^{(j)}}{n_{k+1}^{(j)}-n_k^{(j)}},
\end{eqnarray}
where $k$ is the $k-th$ burst and $n_k$ is the time in which the $k-th$ burst started. The phase synchronization can be found through the Kuramoto's order parameter,
\begin{eqnarray}
r_n= \frac{1}{N}\left| \sum_{j=1}^{N}\exp^{i\varphi_n^{(j)}} \right|,
\end{eqnarray}
when this value is one, it means the network is fully synchronized in phase and when this value is zero, it means the network is fully desynchronized \cite{Kuramoto_75}. It is known that neuronal networks described by Eqs. (\ref{eq1}) and (\ref{eq2}) exhibit phase synchronization when the coupling strength is increased up to a certain value \cite{Batista_1}, as we will show in the next section. 

The second step in our analysis is to study how lesions spreading in the network cause desynchronization. To study this fact, we will assume two different types of lesions:

\begin{description}
\item[Type 1.] Lesions that disrupt the connection between the neurons.
\item[Type 2.] Lesions that kill neurons.
\end{description}

\begin{figure}[t]
\begin{center}
\includegraphics[width=1.0\columnwidth]{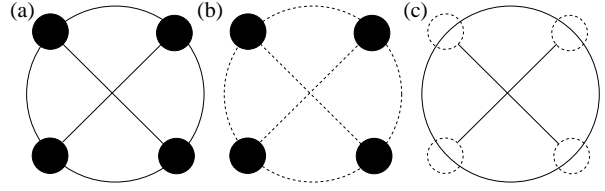}
\end{center}
\caption{Network representation. Panel (a) shows a fully connected network, panels (b) and (c) show the effect of lesions type 1 and type 2, respectively. The dashed lines show where the damage caused by the lesion type is.}
\label{fig1}
\end{figure}

In Fig. \ref{fig1} (a), we show a representation for a network of globally coupled neurons; the effect of lesions type 1 are represented in sout{the Figure} Fig. \ref{fig1} (b), while the effect for lesions type 2 are shown in Fig. \ref{fig1} (c).  

We also consider that for each type of lesion, the coupling strength can be affected by three different situations:

\begin{description}
\item[Reinforced coupling.] For every new damaged neuron, the coupling strength is increased by $\varepsilon(t)=\varepsilon_0/(N-N_d)$. 
\item[Invariant coupling.] The coupling strength does not change with the lesion size, so $\varepsilon(t)=\varepsilon_0/N$.
\item[Reduced coupling.] For every new damaged neuron, the coupling strength is decreased by $\varepsilon(t)=\varepsilon_0/(N+N_d)$.
\end{description}
Here, $N_d$ is the number of disconnected neurons and $\varepsilon_0$ is the initial coupling strength.

\section{Results and Discussion}

\begin{figure}[t]
\begin{center}
\includegraphics[width=1.0\columnwidth,trim=0 1.5cm 0 1.5cm,clip]{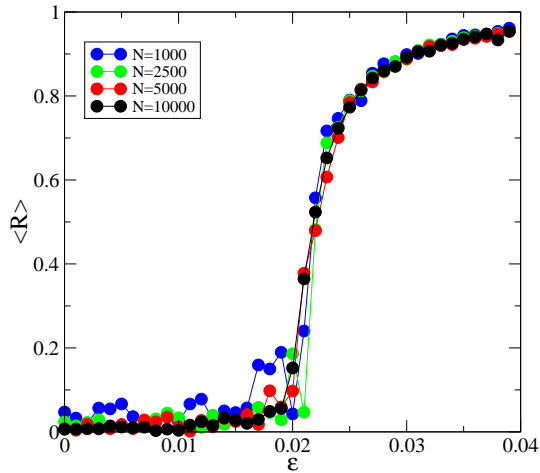}
\end{center}
\caption{The mean order parameter as a function of the coupling strength. The different colors represent different network sizes. Here, $\langle R \rangle$ is the order parameter averaged over the whole network for a time series of 10000 discrete steps after 80000 transient times.}
\label{fig2}
\end{figure}

The first step is to find the values for the coupling strength in which the network shows phase synchronization. In Fig. \ref{fig2}, we show that when the coupling strength is below $\varepsilon_c=0.02$ then the system is completely desynchronized (disregarding fluctuations $\sim 1/\sqrt{N}$). Above this critical value, the order parameter increases, and close to $\varepsilon=0.04$, the network can be considered fully synchronized. Based on that for our results, we will use as initial coupling strength $\varepsilon_0=0.04$. From Fig. \ref{fig2}, we can also see that the transition toward synchronization is invariant with respect to the network size.

\begin{figure*}[t]
\begin{center}
\includegraphics[width=1.5\columnwidth,clip]{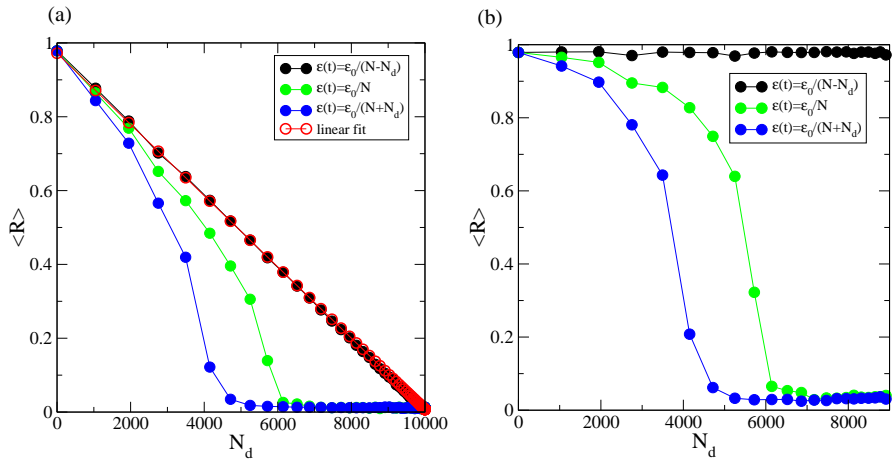}
\end{center}
\caption{The desynchronization process induced by lesions. Panel (a): lesions type 1, panel (b): lesions type 2. The different colors indicate the three different coupling effects: reinforced, invariant and reduced (black, green and blue, respectively). Here, $N_d$ is the number of affected neurons (disrupted for (a) and killed for (b)) and $N=10000$.}
\label{fig3}
\end{figure*}

For lesions type 1, when $\varepsilon(t)$ is reinforced after each new damaged neuron, the synchronization (characterized by the mean order parameter) decays linearly with the number of disconnected neurons ($N_{d}$) but the network just becomes completely desynchronized when all the neurons are lesioned, see Fig. \ref{fig3} (a). For the cases where $\varepsilon(t)$ is invariant or reduced, we observe a roughly first order phase transition where the order parameter decreases linearly up to a critical size of lesioned neurons $N_{d, critical}$ and the whole network desynchronizes. This fact is absent for the reinforced case because increasing the coupling strength increases $N_{d, critical}$ such that the first order phase transition is never observed. The three color lines in Fig. \ref{fig3} (a) indicate that the smaller the coupling strength becomes, the faster the complete phase desynchronization happens. 

An interesting effect occurs for lesions type 2, shown in Fig. \ref{fig3} (b). When the coupling strength is reinforced after each lesion, we do not observe desynchronization. This phenomenon is caused by the fact that when neurons die they do not contribute to the global effect in the network and the remaining neurons being more strongly connected remain synchronized. This fact is not observed for the cases in which the coupling strength remains the same (green line) or is reduced (blue line). For lesions type 2, when the coupling strength decreases toward $N_{d, critical}$, the observed decay follows a roughly second order phase transition, see Fig. \ref{fig3} (b).

\section{Conclusions}

Here, we have investigated two types of lesions and their effects. The presence of phase transition from synchronized to desynchronized state was observed for all cases except for lesions type 2 when the network is reinforced. We have observed that lesions type 1 obey a roughly first order phase transition while lesions type 2 obey a roughly second order phase transition. For both types of lesions, when the coupling strength is continuously reduced after each new damage, then the network desynchronization is faster. Based on that, increasing the coupling strength can be a strategy to compensate the desynchronization effect induced by lesions, but this strategy is more effective for lesions type 2.

The two distinct phase transitions allow us to define a characterization scheme: if the synchronization decays linearly, we could say that we are dealing with a lesion type 1 while if the synchronization decays non-linearly, then a lesion type 2 could be the case. However, a mixture of events could also be observed and then a characterization would become difficult to achieve. Despite this fact, our results show that even simplified models are useful to understand and classify types of lesions and advances in this segment could be helpful to understand the progress of neurodegenerative diseases.

\begin{acknowledgements}
This work has financial support from the Brazilian research agencies CNPq and CAPES. We acknowledge Carlos A. S. Batista for relevant discussions. 
\end{acknowledgements}

\end{document}